\documentclass[12pt,preprint]{aastex}

\begin{document}

\title{Triggering Collapse of the Presolar Dense Cloud Core 
and Injecting Short-Lived Radioisotopes with a Shock Wave. 
IV. Effects of Rotational Axis Orientation}

\author{Alan P.~Boss and Sandra A.~Keiser}
\affil{Department of Terrestrial Magnetism, Carnegie Institution for
Science, 5241 Broad Branch Road, NW, Washington, DC 20015-1305}
\email{boss@dtm.ciw.edu}

\begin{abstract}

Both astronomical observations of the interaction of Type II supernova 
remnants (SNR) with dense interstellar clouds as well as cosmochemical 
studies of the abundances of daughter products of short-lived radioisotopes 
(SLRIs) formed by supernova nucleosynthesis support the hypothesis that 
the Solar System’s SLRIs may have been derived from a supernova. This paper 
continues a series devoted to examining whether such a shock wave could 
have triggered the dynamical collapse of a dense, presolar cloud core 
and simultaneously injected sufficient abundances of SLRIs to explain 
the cosmochemical evidence. Here we examine the effects of shock waves 
striking clouds whose spin axes are oriented perpendicular, rather than 
parallel, to the direction of propagation of the shock front. The models 
start with 2.2 $M_\odot$ cloud cores and shock speeds of 20 or 40 km 
s$^{-1}$. 
Central protostars and protoplanetary disks form in all models, though with 
disk spin axes aligned somewhat randomly. The disks derive most of their 
angular momentum not from the initial cloud rotation, but from the 
Rayleigh-Taylor fingers that also inject shock wave SLRIs. Injection 
efficiencies, $f_i$, the fraction of the incident shock wave material 
injected into the collapsing cloud core, are ∼ 0.04 - 0.1 in these models, 
similar to when the rotation axis is parallel to the shock propagation 
direction. Evidently altering the rotation axis orientation has only a 
minor effect on the outcome, strengthening the case for this scenario 
as an explanation for the Solar System’s SLRIs.

\end{abstract}


\section{Introduction}

 Radio and sub-millimeter observations of the Type II supernova (SNe) remnant 
(SNR) W44 and its interaction with the W44 giant molecular cloud complex 
imply the presence of dense clumps of molecular gas that have been struck 
and compressed to sizes much less than 0.3 pc by the SNR shock front, with
shock speeds of 20 to 30 km s$^{-1}$ (Sashida et al. 2013; Reach et al. 2005).
HST images of the Cygnus Loop Type II SNR reveal shock front thicknesses 
of order $2 \times 10^{-4}$ pc (Blair et al. 1999). Detailed two dimensional
(2D) axisymmetric (Boss \& Keiser 2013) and fully three dimensional (3D) 
(Boss \& Keiser 2012, 2014; Li et al. 2014)
hydrodynamical models of shock waves with these speeds and
thicknesses have demonstrated the viability of triggered collapse of dense 
cloud cores with simultaneous injection of shock front material into the
resulting collapsing protostar and protoplanetary disk. Given the problems
with shocks waves associated with AGB and WR star winds (Boss \& Keiser 2013), 
a SNR shock appears to be the leading contender for achieving 
simultaneous triggering and injection of core-collapse supernova-derived
short-lived radioisotopes (SLRIs, e.g., $^{60}$Fe and $^{26}$Al) into the presolar 
cloud and the resulting solar nebula (Cameron \& Truran 1977; Boss 1995).

 Cosmochemical evidence for live $^{60}$Fe during the formation of refractory 
inclusions and chondrites has waxed and waned in recent years. Data on 
ferromagnesian chondrules from two ordinary chondrites (OC) implied initial 
$^{60}$Fe/$^{56}$Fe ratios of $\sim 5 - 10 \times 10^{-7}$ (Tachibana et al. 
2006), whereas bulk sample data from a wide range of meteorites suggested 
an initial ratio of only $\sim 1.15 \times 10^{-8}$ (Tang \& Dauphas 2012).
A combined study of $^{60}$Fe and $^{26}$Al in chondrules from unequilibrated 
OC (UOC) implied an initial $^{60}$Fe/$^{56}$Fe ratio of $\sim 7 \times 10^{-7}$ 
(Mishra \& Goswami 2014) and supported a SNe as the source of the SLRIs. Analyses 
of other chondrules from UOC yielded ratios in the range of 
$\sim 2 - 8 \times 10^{-7}$ (Mishra \& Chaussidon 2014). While there
is as yet no explanation for the discrepancy between bulk samples and 
chondrule fragments (Mishra \& Chaussidon 2014), 
a SNe remains as a plausible source for the SLRIs. Indeed, evidence
for possible live $^{135}$Cs in CAIs suggests its origin in a Type II SNe (or
its WR progenitor) close to the presolar cloud (Bermingham et al. 2014). 
Furthermore, the evidence for live $^{10}$Be in FUN-CAIs appears consistent 
with $^{10}$Be formation in the presolar molecular cloud by galactic 
cosmic rays (GCRs) emitted by a SNR that thereafter triggered the 
collapse of the presolar cloud core (Tatischeff et al. 2014). Statistical analysis
of the SLRI enrichments expected for disks and cloud cores in clusters
containing massive stars showed that cloud cores receive a larger dose on average
than disks (Adams et al. 2014), favoring the presolar cloud scenario 
(Cameron \& Truran 1977; Boss 1995) over the alternative mechanism of injection 
into a previously formed solar nebula (Ouellette et al. 2007, 2010).

 Given this strong support for the triggering and injection
hypothesis for the origin of the SLRIs, we continue to calculate increasingly
detailed models of the hydrodynamics of the shock-cloud interactions, in order to
further understand the physics of the process. Our previous work
in both 2D (Boss \& Keiser 2013) and 3D (Boss \& Keiser 2014)
considered rotating presolar clouds, capable of collapsing
to form protostellar disks. Because of symmetry constraints in 2D, these models
assumed that the cloud rotation axis was {\it parallel} to the direction 
of shock wave propagation. For ease of comparison, our first 3D rotating 
cloud models (Boss \& Keiser 2014) made this same 
assumption. Here we present the results of a set of 3D models identical to
previous rotating cloud models, but with the rotation axis {\it perpendicular} 
to the shock front direction. We seek to discover what effect, if any, the
rotational axis orientation might have on the SNe trigger and injection 
hypothesis.

\section{Numerical Hydrodynamics Code}

 The numerical models were calculated with the FLASH2.5 adaptive mesh
refinement (AMR) hydrodynamics code (Fryxell et al. 2000) in exactly the
same manner as in our previous 3D models (Boss \& Keiser 2012, 2014), except
for changing the initial direction of the target cloud rotation axis to
be perpendicular to the shock propagation direction. We give here a brief
summary of the basic details of our implementation of the FLASH2.5 AMR code; 
further details about our implementation may be found in Boss et al. (2010).
 
 We employed FLASH2.5 in Cartesian coordinates ($x, y, z$). FLASH uses a 
block-structured adaptive grid approach, using the PARAMESH package, to enhance
the grid resolution in regions with strong gradients of chosen variables.
Initially the numerical 
grid has six grid blocks along the $\hat x$ and $\hat z$ axes and nine along the 
$\hat y$ axis. The $-\hat y$ axis is the direction of propagation of the shock wave, 
while the rotation axis of the target cloud is the $\hat x$ axis. 
The grid blocks each consist of $8^3$ grid points with four levels 
of grid refinement initially. Refined grids have one half the grid spacings of 
their parent grids. We use both the density and the color field (defined
below) to determine when grid refinement is necessary to follow the shocked
regions. While Boss et al. (2010) found that adequate resolution was obtained
with a maximum of five or six levels of grid refinement, here the maximum 
number of levels of grid refinement permitted was increased in some models
to seven levels during the evolutions, when the memory allocation of the flash 
cluster nodes used for the calculations permitted the use of yet another 
level of grid refinement. Additional levels of grid refinement (beyond four) 
are necessary in order to attain the highest possible resolution of the dynamically 
collapsing regions that occur late in the evolutions. With seven levels, 
the smallest grid spacing reached is then $\sim 4 \times 10^{-5}$ pc $\sim 9$ AU. 

 As in our previous models (Boss et al. 2010; Boss \& Keiser 2012, 2013, 2014),
we used a color field to follow the evolution of shock front gas and dust.
This allows us to quantify the efficiency with which the shock front matter
is injected into the target molecular cloud core. The target molecular cloud core 
and the surrounding, ambient low density gas are initially isothermal at 10 K. 
The shock front and post-shock gas are initially isothermal at 1000 K. 
This initial shock temperature follows from the C-shock models of
Kaufman \& Neufeld (1996), who found peak shock temperatures of $\sim 10^3$K
for shock speeds in the range of 5-45 km s$^{-1}$ (see Boss et al. 2010).
We then follow the compressional heating as well as cooling of the 
interacting shock-cloud system by molecular species. In particular, we
assume cooling at a rate consistent with cooling by optically thin H$_2$O, 
CO, and H$_2$ molecules in order to follow the thermodynamics of the 
shock-cloud interaction. As before, we used Neufeld \& Kaufman's 
(1993) radiative cooling rate of $\Lambda \approx 9 \times 10^{19} (T/100) 
\rho^2$ erg cm$^{-3}$ s$^{-1}$, where $T$ is the gas-dust temperature in K 
and $\rho$ is the gas density in g cm$^{-3}$. This rate is appropriate for 
cooling caused by rotational and vibrational transitions of optically thin, 
warm molecular gas (Boss et al. 2010). Note that we do not try to follow
the collapsing protostellar objects and disks deep into the nonisothermal collapse
regime, after the collapsing central regions become optically thick at densities above
$\sim 10^{-12}$ g cm$^{-3}$; we leave this additional physical complication
for our future models.

\section{Initial Conditions}

 Table 1 lists the variations in the initial conditions for the six new 3D models. 
These models have all been previously calculated with the rotational axis aligned 
with the shock propagation direction, allowing a direct comparison of the results in
each case (Boss \& Keiser 2014). All models start with 2.2 $M_\odot$ molecular
cloud cores with radii of 0.053 pc and Bonnor-Ebert radial density profiles.
The clouds have solid body rotation about the $\hat x$ axis with an angular
frequency of either $\Omega_c = 10^{-14}$ or $10^{-13}$ rad s$^{-1}$. These angular 
frequencies are typical of observed dense cloud cores (e.g., Harsono et al. 2014).

 The shock waves propagate toward the $- \hat y$ direction with shock speeds of 
$v_s = 20$ or 40 km s$^{-1}$. The shock widths are $3 \times 10^{-4}$ pc for all 
the models, i.e., the shock widths $w_s$ = 0.1 in units of the standard shock width, 
0.0030 pc (Boss et al. 2010). The initial shock densities ($\rho_s$) are 
$7.2 \times 10^{-18}$ g cm$^{-3}$ or $1.44 \times 10^{-17}$ g cm$^{-3}$, 
i.e., 200 or 400 times the reference shock density of $3.6 \times 10^{-20}$ 
g cm$^{-3}$. Details about why these particular shock wave parameters
are appropriate for a supernova shock that has travelled several pc through
the interstellar medium prior to smacking the target clouds may be found
in Boss \& Keiser (2012, 2013, 2014). In particular, Boss et al. (2010)
studied the effects of shock speeds ranging from 1 km s$^{-1}$ to 100 km s$^{-1}$,
and found that only shock speeds in the range of 5 km s$^{-1}$ to 70 km s$^{-1}$
were able to trigger collapse and injection into the target cloud assumed here.
Boss \& Keiser (2013) studied the effects of different shock widths ($w_s$ = 0.1,
1.0, and 10.0). Our use of $w_s$ = 0.1 is consistent with observations of
the Cygnus Loop (Blair et al. 1999) and other SNRs (see Boss \& Keiser 2014). 

\section{Results}

 As before (Boss \& Keiser 2012, 2013, 2014), the shocks strike
the top edge of the target clouds and compress the central
regions into sustained self-gravitational collapse, while accelerating
the clouds downstream in the shock direction. The molecular line cooling
keeps the temperatures of the shocked region below 1000 K, while the 
un-shocked regions remain isothermal at 10 K. Rayleigh-Taylor (R-T) fingers
form at the shock-cloud interface, and these R-T fingers are directly
responsible for injecting supernova-derived material into the collapsing
cloud core. The calculations are terminated once the maximum densities
exceed $\sim 10^{-11}$ g cm$^{-3}$, as at those densities the collapsing
regions have become optically thick and will no longer be able to cool
at the rate appropriate for optically thin regions.

 Figures 1 through 6 show the basic results for the six models defined in 
Table 1. It is evident that central protostars and edge-on disks form in 
all six models, though with overall disk structures considerably different
than in models where the shock direction is parallel to the cloud
rotation axes (Boss \& Keiser 2014). In the latter case, the edge-on
disks are always aligned with their innermost midplanes perpendicular to
the shock direction, i.e., with their rotational axes aligned with the
shock direction (cf., Figure 4 of Boss \& Keiser 2014). This is a
direct consequence of their initial rotational axes pointing downward, in 
the sense of Figures 1 through 6. The outermost disks, however, in the
aligned axes models do warp upwards or downwards compared to the
innermost disk midplane, beyond distances of $\sim 150$ AU for
$\Omega_c = 10^{-14}$ rad s$^{-1}$ and beyond $\sim 500$ AU
for $\Omega_c = 10^{-13}$ rad s$^{-1}$. In comparison, the edge-on 
disks seen in Figures 1 through 6 in some cases have innermost disk 
midplanes tilted compared to the perpendicular to the shock direction 
(e.g., Figures 3, 4, 5), while the remainder are closer to
being perpendicular (Figure 1, 2, and 6). In all six models, however,
the outermost disks warp away from the midplane orientations.

 What is most remarkable about the disk orientations in Figures 1 through 
6 is the fact that in no case are the disk midplanes seen to be aligned
with the direction of the shock propagation, i.e., the $- \hat y$ axis,
as might be expected considering that the initial sense of rotation of 
the target clouds is around an axis aligned with the $+ \hat x$ axis.
This is quite different from the results for the Boss \& Keiser (2014)
shock-aligned axes models, where the disk midplanes are perpendicular
to the initial cloud spin axes, implying that the disks have derived their
angular momentum from the initial target clouds. Given the fact that 
the new disks are warped and otherwise highly distorted (see 
especially Figure 5), it is unclear how to assign a single, specific 
direction of an overall disk rotational axis, as there evidently is 
no single rotational axis for these models.  {\bf Considering the situation 
evident in Figures 1 through 6, Table 1 lists the variety of spin 
directions ascertained for different portions of each disk.}

 The reason for this puzzling difference in the results can be
discovered first by examining the spin directions of the material in the
disks. Figure 7 shows the $z$ component of the velocity field for 
model 40-200-pr13, plotted in the same sense as in Figure 1 for this
model. Figure 7 shows that the axis of rotation of the innermost disk 
is along the $+ \hat y$ axis, in the opposite direction of the shock
propagation, and perpendicular to the initial target cloud rotation
axis. Figure 8 shows the same type of plot for model 20-400-pr14,
where the axis of rotation is now aligned with the $- \hat y$ axis,
the opposite direction of model 40-200-pr13, again perpendicular to the 
initial target cloud rotation axis. Table 1 summarizes the
approximate disk spin axis directions for all six models, showing
no obvious pattern for the overall orientations. Evidently the disks
seen in Figures 1 through 6 did not derive the majority of their
angular momentum from the assumed initial cloud angular momentum.

 Further investigation of Figures 7 and 8 reveals the source of the 
bulk of the angular momentum contained in the disks. Figure 7 shows that 
the largest variations in the $z$ velocity field are found in the R-T
fingers in the shock front. In fact, in Figure 7, a clear stream of
positive $v_z$ gas can be traced from the R-T fingers down to the
innermost disk midplane. Similarly, in Figure 8 a large field of
negative $v_z$ gas connects the disk midplane with the R-T fingers.
This explanation is made even more robust in Figures 9 and 10,
where the disk formed in model 40-200-pr13 is seen face-on, with a 
strong trailing one-arm spiral indicative of the disk spinning with its
rotational axis aligned with the $+ \hat y$ direction. Figure 10
plots the $v_x$ velocity component for this model's midplane,
showing the same sense of rotation as previously inferred. A plot of
$v_z$ shows the same agreement. Again, Figure 10 shows that the
greatest variations in $v_x$ originate in the R-T fingers, and in 
fact streams of similar velocity gas in the disk can be traced
back to R-T fingers with the same velocity sign. Evidently the R-T
fingers provide the bulk of the angular momentum for disks formed
when the initial rotational axes are not aligned with the shock
propagation direction. In general, one would not expect these two
to be aligned, so it is important to understand how this could happen.

 To further investigate this odd result, it is important to note that when 
the initial 3D cloud is not rotating at all, i.e., when $\Omega_c = 0$, 
no obvious disks form at all, in spite of R-T fingers being present
(Boss \& Keiser 2012). Instead, the collapsing protostar is roughly
spherical (cf., Figure 3 in Boss \& Keiser 2012). This implies that
it is the interaction of the initial target cloud spin with the R-T
fingers formed at the shock-cloud interface that leads to the R-T
fingers being sheared and deflected away from motions that would
otherwise be more purely directed downward, along the $- \hat y$ axis.
I.e., the $v_y < 0$ momentum of the shock gets converted by the
sheared R-T fingers into a complex pattern of $v_x$ and $v_z$ momentum,
sufficient to provide the angular momentum of the disks seen in Figures
1 through 6. In support of this explanation, note that the 
Keplerian velocity of gas orbiting at 100 AU around a protostar
with a mass of 0.5 $M_\odot$ is $\sim$ 2 km $^{-1}$. Figure 7 shows 
that $\sim$ 2 km $^{-1}$ is indeed a typical speed for the $z$ velocity 
component of the gas 100 AU out in the rotating edge-on disk in
model 40-200-pr13, consistent with a disk in Keplerian rotation.
Similarly, Figure 10 illustrates exactly the same situation
for the $x$ component of the gas velocity at 100 AU from the central
protostar in the face-on disk seen in Figure 9, for this same model. 
These speeds are the same as can be seen in the R-T fingers evident in 
Figures 7 and 10, from which the disk gas originated, and cement the 
case for the angular momentum sufficient to form rotating, Keplerian
protostellar disks being derived from the R-T fingers, rather than from
the initial cloud rotation. 

 In this same context, it is important to note that the shocks start with 
considerably higher speeds along $\hat y$ (20 to 40 km s$^{-1}$) compared to 
the initial target cloud rotational speeds (no more than $\sim$ 0.1 km s$^{-1}$  
for $\Omega_c = 10^{-13}$ rad s$^{-1}$, and no more than $\sim$ 0.01 km s$^{-1}$ 
for $\Omega_c = 10^{-14}$ rad s$^{-1}$), so it is easy to see how even a small
deflection of the shock front momentum carried by the R-T fingers can 
dominate the outcome for the disks that form. This phenomenon also
explains the origin of the outer disk warps seen in the models with
aligned rotational axes and shock directions (Boss \& Keiser 2012):
the addition of material to the disk with velocities influenced
by the R-T fingers.

 This explanation appears to be consistent with the quite different 
result of Li et al. (2014), who showed one model (R2) where an initial 
cloud rotating similar to the present models produced a disk with a
spin axis in the expected direction (cf., Figure 2(d) in Li et al. 2014).
While there are many differences between the present models and those
of Li et al. (2014), e.g., their use of sink cells and our use of detailed
thermodynamics, the fact that their R2 model started with a shock
speed of 3 km s$^{-1}$ implies that the shock-derived momentum was unable 
to overcome the initial cloud rotational motions. In addition, R-T fingers
do not appear in their calculations, apparently because of insufficient
spatial resolution, in spite of their using an AMR code, ASTROBEAR 2.0
(Li et al. 2014). However, Li et al. (2014) used only three levels
of grid refinement, leading to a smallest grid size of about 23 AU,
whereas the present models with seven levels of refinement had a smallest
grid size of about 9 AU, which could be an important factor. Combined 
with the other differences between these two sets of models, the fact 
that Li et al. (2014) did not observe R-T fingers in any of their
models ensures that, unlike the present models, where R-T fingers are
essential to the injection process, R-T fingers could not dominate the
angular momentum of the disk that formed in the Li et al. (2014) model 
R2.

 These calculations imply that besides injecting the SLRIs found in
primitive meteorites, R-T fingers may have also delivered the angular
momentum that allowed a protostellar disk to form around the nascent
protosun, a disk that would evolve to become the protoplanetary disk
that formed our Solar System. The fact that the disk spin axis 
orientations listed in Table 1 are somewhat random implies that
protoplanetary disks formed by this mechanism would be expected to
have spin axes with roughly random orientations, orientations that are
determined by the detailed development of the R-T fingers that delivered 
the disks' angular momenta. 

 We now turn to a consideration of the injection efficiencies for these
models. As before (Boss \& Keiser 2012, 2013, 2014), the injection 
efficiency $f_i$ is defined as the fraction of the incident shock wave 
material that is injected into the collapsing cloud core. Table 1
lists the calculated values of $f_i$, along with the mass ($M_c$) of the 
dynamically collapsing regions (i.e., regions with $\rho > 10^{-16}$ 
g cm$^{-3}$) and the mass of the shock wave-derived material in this 
same region ($M_s$). The calculated values of $f_i$ compare favorably
with those for otherwise identical models with aligned rotation
(cf., Table 1 in Boss \& Keiser 2014). E.g., model 40-200-pr13 has
$f_i$ = 0.037, whereas the corresponding aligned rotation model from
Boss \& Keiser (2014) had $f_i$ = 0.032. This same minor increase in $f_i$
is seen for the other models in Table 1 with corresponding aligned axes
models as well. Clearly having non-aligned axes does not make SLRI injection
any harder. The increased values of $f_i$ can be attributed to the larger
role of the R-T fingers in the non-aligned models in providing the bulk
of the disk angular momentum, along with the SLRIs. Figures 11 and 12 show
the color fields for models 40-200-pr13 and 20-400-pr14, making it clear
that by this advanced phase of the evolutions, the R-T fingers have
successfully injected shock wave material throughout the collapsing
regions.

 While the new models exhibit the robustness of our previous estimates of
$f_i$, it must also be noted that the SNe shock waves must sweep up
considerable amounts of intervening interstellar gas and dust in order
to slow down to the shock speeds considered here and previously found
to be the most favorable for injection (e.g., Boss et al. 2010; Boss 
\& Keiser 2013). This implies a large amount of dilution of the pristine
SLRIs formed by the SNe. As in Boss \& Keiser (2012), one can define a 
factor $\beta$ to be the ratio of the shock mass originating in the SNe to 
the mass swept up in the intervening gas and dust. For the present shock 
speeds, a value of $\beta \sim 0.01$ is appropriate (Boss \& Keiser 2012).
The dilution factor $D$ is the ratio of the amount of mass derived from 
the supernova to the amount of mass derived from the target cloud. For 
models with $\rho_s = 200$ (Table 1), the amount of mass in the incident
shock wave is 0.30 $M_\odot$. Assuming a final mass for the protostar and
disk of 1.0 $M_\odot$, this results in $D \sim 3 \times 10^{-4}$. This
dilution factor is in reasonable agreement with estimates for SNe-derived
SLRIs of $D \sim 10^{-4}$ to $D \sim 10^{-3}$ (Takigawa et al. 2008). In spite
of the ongoing uncertainty over the correct initial abundances of SLRIs
in the solar nebula, as discussed in the Introduction, the dilution factor
estimates of Takigawa et al. (2008) appear to be the best current values,
as argued in Boss \& Keiser (2014). {\bf Li et al. (2014) found insufficient 
SLRI injection in their models to explain the cosmochemical evidence. 
However, as noted above, Li et al. (2014) were not able to resolve 
the R-T fingers necessary for significant injection, and furthermore studied 
only shock speeds of 3 or 6 km s$^{-1}$, at or below the lower end of the 
range (5 to 70 km s$^{-1}$) found necessary for simultaneous triggering 
and injection by Boss et al. (2010).}

 Finally, we note from Table 1 that the injection efficiencies $f_i$ 
can be seen to increase as the amount of mass $M_c$ in the collapsing
region with $\rho > 10^{-16}$ g cm$^{-3}$ increases. Evidently some of
the estimates of $f_i$ in Table 1 may be lower bounds, in that the
values of $f_i$ are likely to increase when the calculation is taken
farther in time and more mass joins the collapsing regions.
In practice, the calculations are halted when the 
collapsing region becomes too dense for the isothermal approximation
to be valid (i.e., $\rho > 10^{-11}$ g cm$^{-3}$), which is also about
when the collapsing regions reach the bottom edge of the computational
domain. Both of these limitations are being addressed in a new set of 
models currently underway.

\section{Conclusions}

 These models have demonstrated that the degree of alignment between a
rotating molecular cloud core and the direction of propagation of a
shock wave that strikes the cloud core, triggering its self-gravitational
collapse, does not have a major effect on the efficiency of injection
of shock-derived SLRIs into the collapsing cloud core. In fact, the
injection efficiencies in the non-aligned models ($\sim$ 0.04 to 0.1)
are somewhat higher than in the aligned models with identical 
shock wave parameters (cf., Boss \& Keiser 2012, 2013, 2014). Hence
from the point of view of the numerical calculations of the mechanism,
the supernovae shock wave triggering and injection hypothesis
remains as a plausible explanation (Cameron \& Truran 1977; Boss 1995).

 Remarkably, these new models have also introduced a new feature of
the shock wave triggering and injection mechanism: the R-T fingers 
responsible for SLRI injection can concomitantly result in the injection
of enough momentum to largely determine the direction of the resulting
disks' spin axis orientations. In such cases, the R-T fingers may have
been responsible not only for the acquisition of the SLRIs inferred to 
have been present in the most primitive meteorites, but also for the
very fact that a rotating protostellar disk was formed, a disk that
eventually led to the formation of our planetary system.
 
 Our future 3D FLASH models will investigate the effects of the loss of
molecular line cooling once the clouds become optically thick at densities 
well above $\sim 10^{-12}$ g cm$^{-3}$. In this regime, the collapsing 
regions begin to heat above 10 K, but continue their collapse toward 
the formation of the first protostellar core, at 
$\rho_{max} \sim 10^{-10}$ g cm$^{-3}$ (e.g., Boss \& Yorke 1995).

\acknowledgments

 We thank the referee for improvements to the manuscript, 
and Michael Acierno for help with the flash cluster
at DTM, where the calculations were performed.
This research was supported in part by NASA Origins of Solar Systems 
grant NNX09AF62G. The software used in this work was in 
part developed by the DOE-supported ASC/Alliances Center for 
Astrophysical Thermonuclear Flashes at the University of Chicago.

\clearpage
\begin{deluxetable}{lccccccc}
\tablecaption{Initial conditions and basic results for the models, with 
varied shock speeds ($v_s$, in units of km s$^{-1}$), shock gas densities ($\rho_s$, 
in units of the standard shock density, $3.6 \times 10^{-20}$ g cm$^{-3}$), target 
cloud perpendicular rotation rates ($\Omega_c$, in rad s$^{-1}$), mass in collapsing 
($\rho > 10^{-16}$ g cm$^{-3}$) region ($M_c$, in in $M_\odot$), shock wave
masses in collapsing regions ($M_s$, in $M_\odot$), injection efficiencies ($f_i$), 
and approximate spin directions of resulting disks. \label{tbl-1}}
\tablewidth{0pt}
\tablehead{\colhead{Model} 
& \colhead{$v_s$}
& \colhead{$\rho_s$} 
& \colhead{$\Omega_c$} 
& \colhead{$M_c$} 
& \colhead{$M_s$} 
& \colhead{$f_i$}
& spin direction }
\startdata

40-200-pr13 & 40 & 200 & $10^{-13}$ & 0.45 & 0.011 & 0.037 & +$\hat y$ \\

40-200-pr14 & 40 & 200 & $10^{-14}$ & 0.49 & 0.012 & 0.040 & -$\hat y$ \\

20-200-pr13 & 20 & 200 & $10^{-13}$ & 0.95 & 0.028 & 0.093 & -$\hat y$ to +$\hat x$ to +$\hat z$ \\

20-200-pr14 & 20 & 200 & $10^{-14}$ & 1.04 & 0.031 & 0.103 & -$\hat y$ to +$\hat x$ to -$\hat z$ \\

20-400-pr13 & 20 & 400 & $10^{-13}$ & 0.64 & 0.032 & 0.053 & -$\hat y$ to -$\hat x$ \\

20-400-pr14 & 20 & 400 & $10^{-14}$ & 0.72 & 0.036 & 0.060 & -$\hat y$ \\

\enddata
\end{deluxetable}

\begin{figure}
\vspace{-1.0in}
\plotone{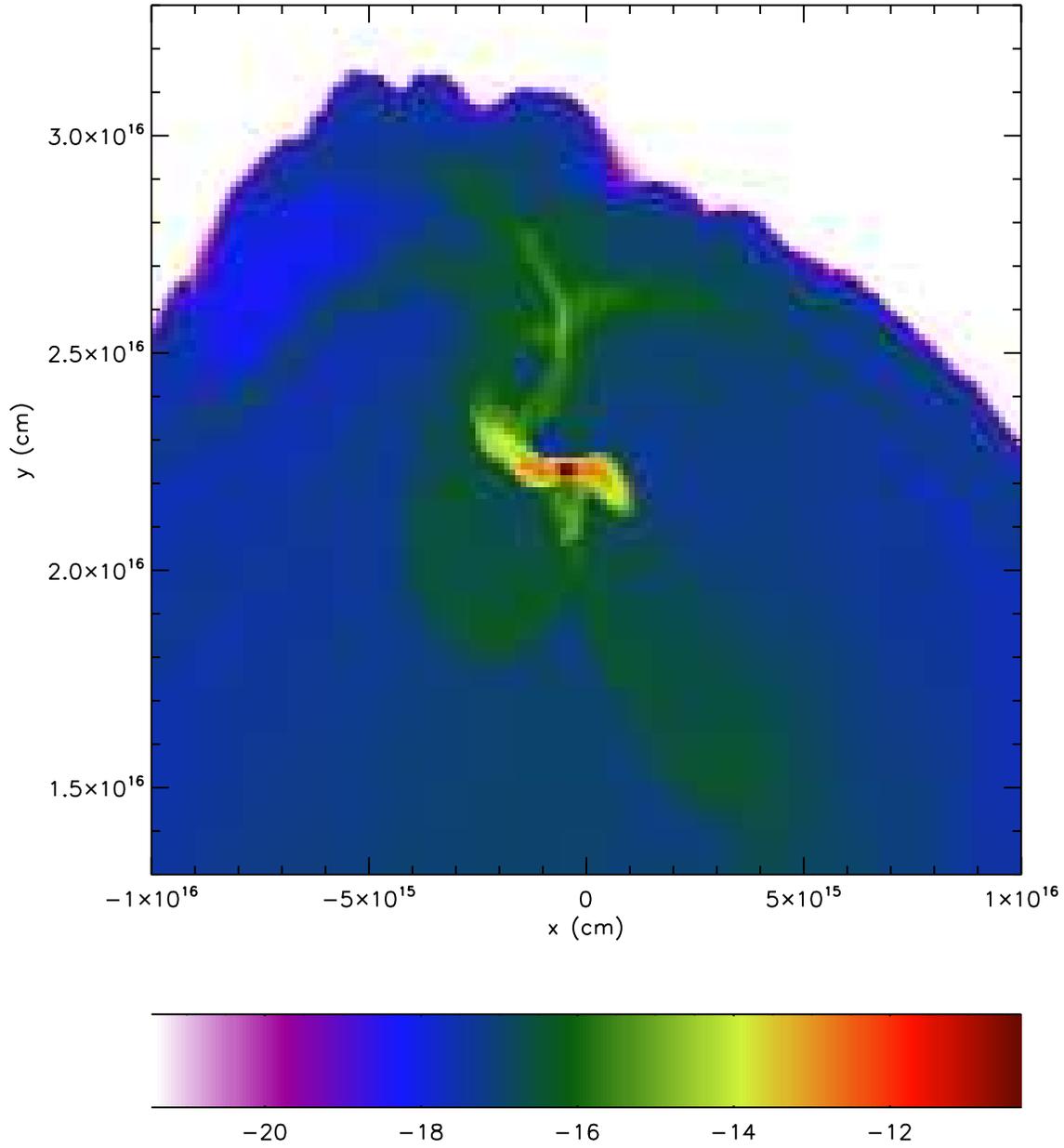}
\vspace{-1.0in}
\caption{Model 40-200-pr13 log density cross-section in the $z = -1.5 \times 10^{15}$ cm 
plane, where the maximum density occurs, showing an edge-on disk. 
Region shown is 1333 AU across at 0.085 Myr. Shock propagates from 
top to bottom in each model. Direction of initial target cloud spin axis is to the right.}
\end{figure}
\clearpage

\begin{figure}
\vspace{-1.0in}
\plotone{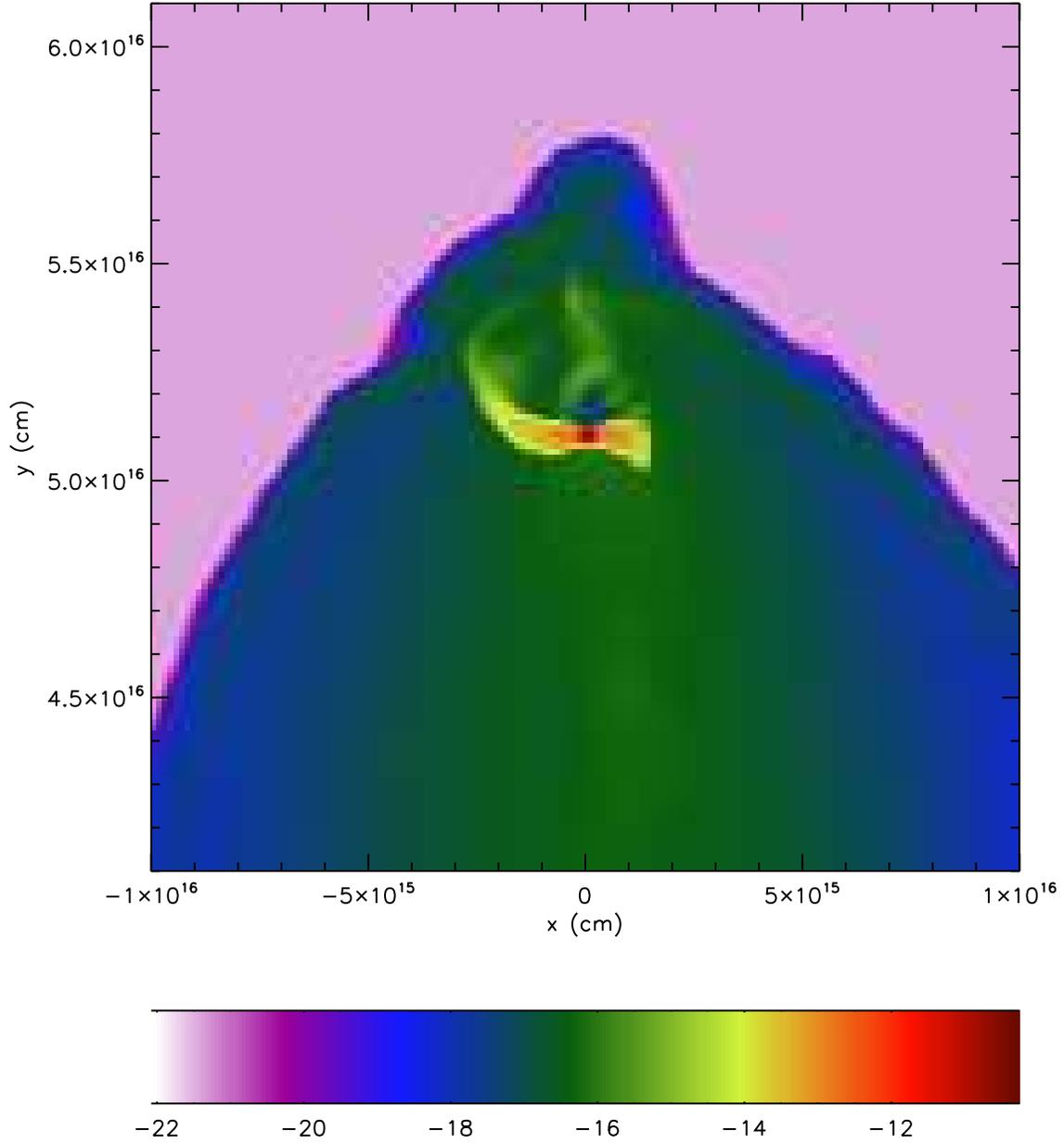}
\vspace{-1.0in}
\caption{Model 40-200-pr14 log density, plotted as in Fig. 1, but 
in the $z = -0.5 \times 10^{15}$ cm plane, at 0.079 Myr.}
\end{figure}
\clearpage

\begin{figure}
\vspace{-1.0in}
\plotone{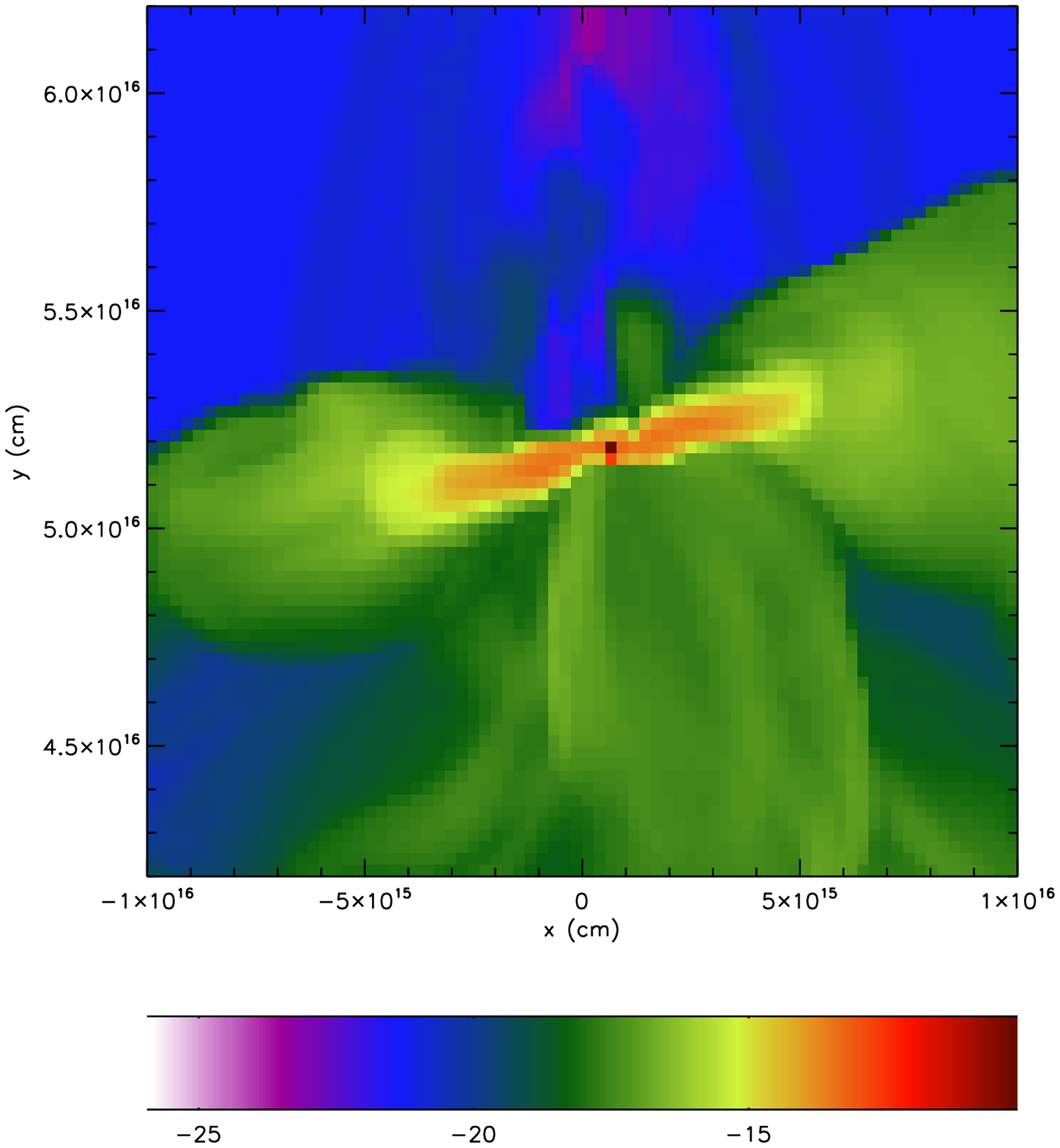}
\vspace{-1.0in}
\caption{Model 20-200-pr13 log density, plotted as in Fig. 1, but
in the $z = -1.8 \times 10^{15}$ cm plane, at 0.15 Myr.}
\end{figure}
\clearpage

\begin{figure}
\vspace{-1.0in}
\plotone{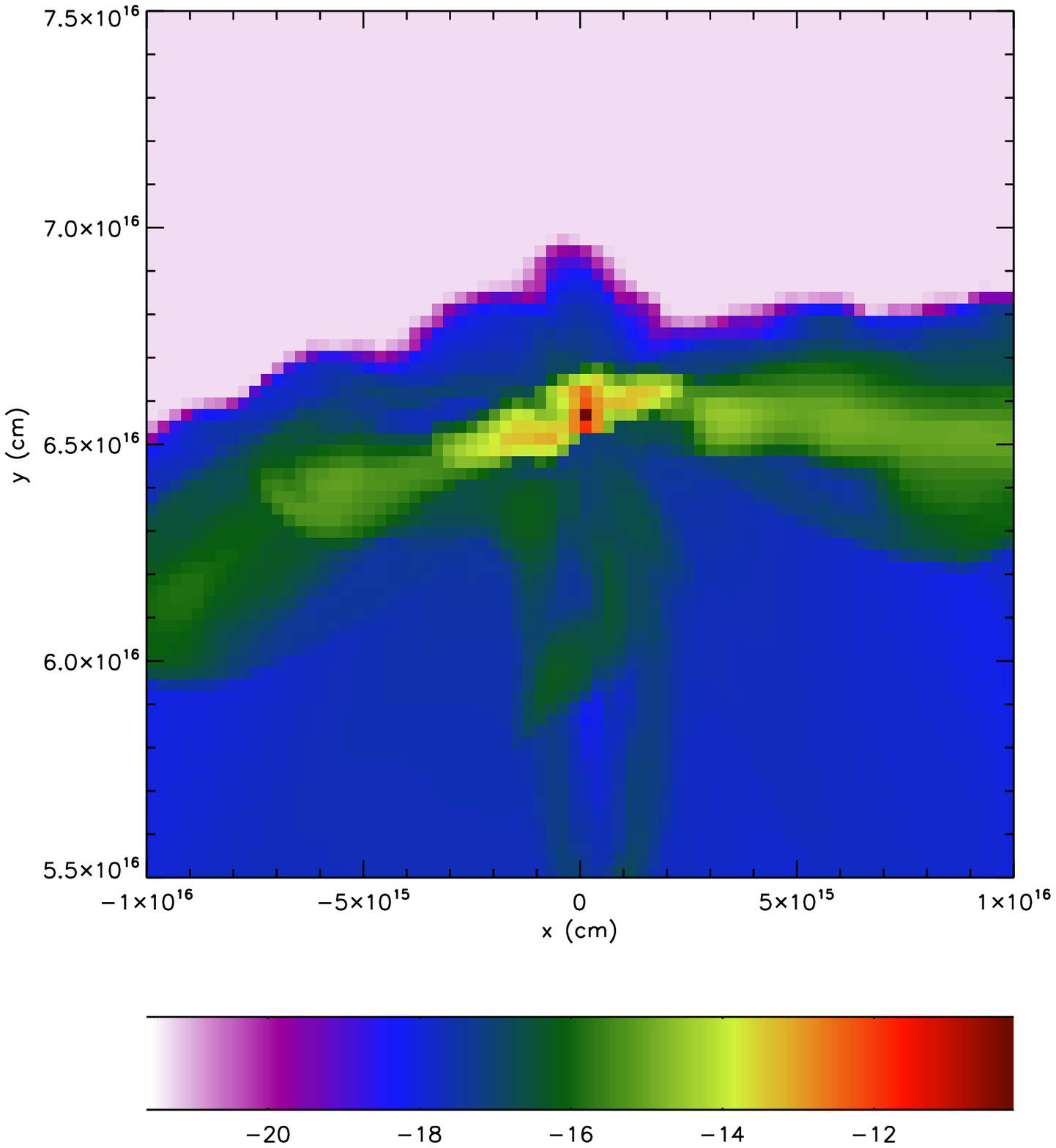}
\vspace{-1.0in}
\caption{Model 20-200-pr14 log density, plotted as in Fig. 1, but
in the $z = 0$ plane, at 0.13 Myr.}
\end{figure}
\clearpage

\begin{figure}
\vspace{-1.0in}
\plotone{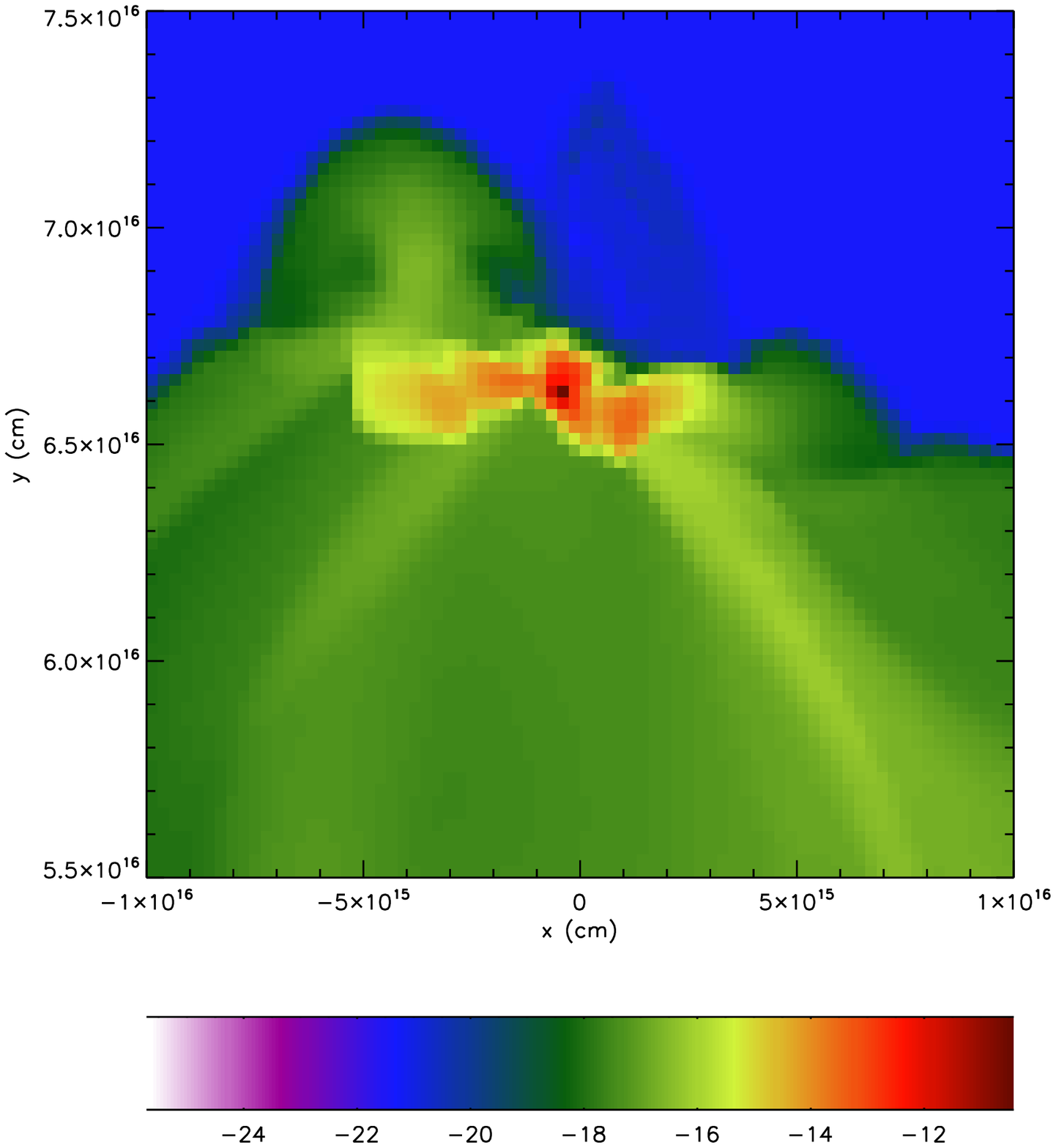}
\vspace{-1.0in}
\caption{Model 20-400-pr13 log density, plotted as in Fig. 1, but
in the $z = -1.2 \times 10^{15}$ cm plane, at 0.093 Myr.}
\end{figure}
\clearpage

\begin{figure}
\vspace{-1.0in}
\plotone{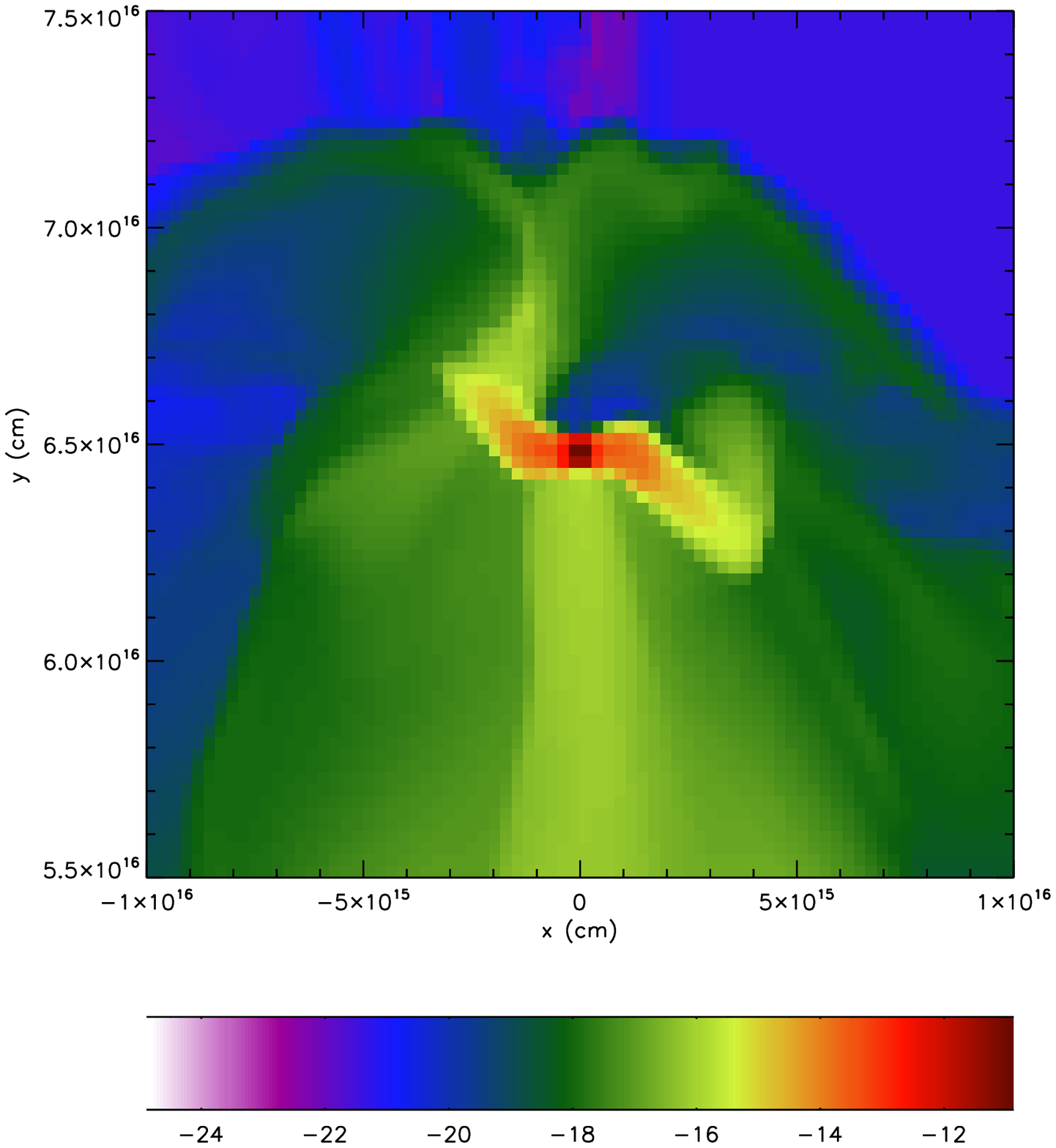}
\vspace{-1.0in}
\caption{Model 20-400-pr14 log density, plotted as in Fig. 1, but
in the $z = 0$ plane, at 0.090 Myr.}
\end{figure}
\clearpage

\begin{figure}
\vspace{-1.0in}
\plotone{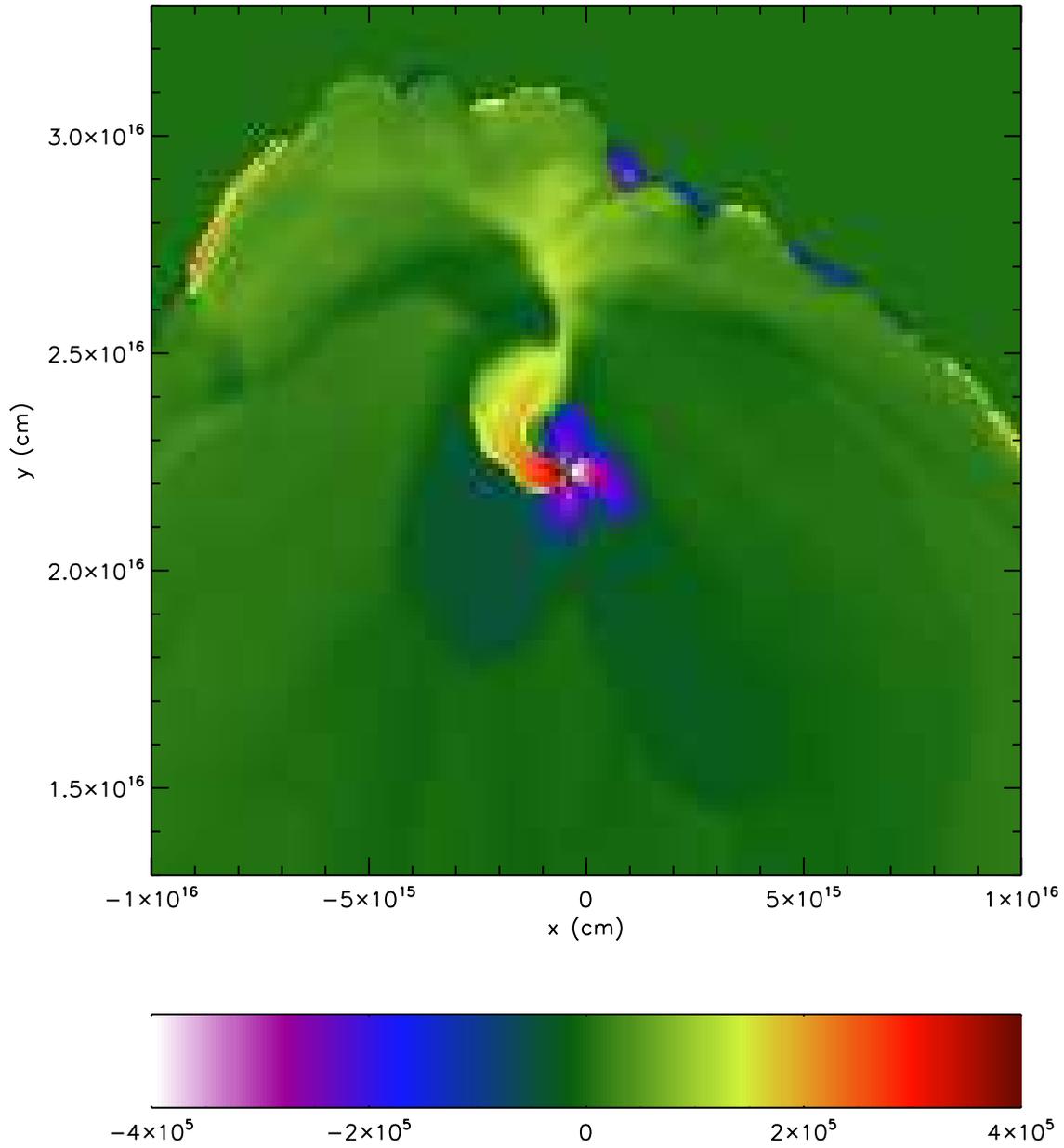}
\vspace{-1.0in}
\caption{Model 40-200-pr13 $z$ velocity, plotted as in Fig. 1 at 0.085 Myr.}
\end{figure}
\clearpage

\begin{figure}
\vspace{-1.0in}
\plotone{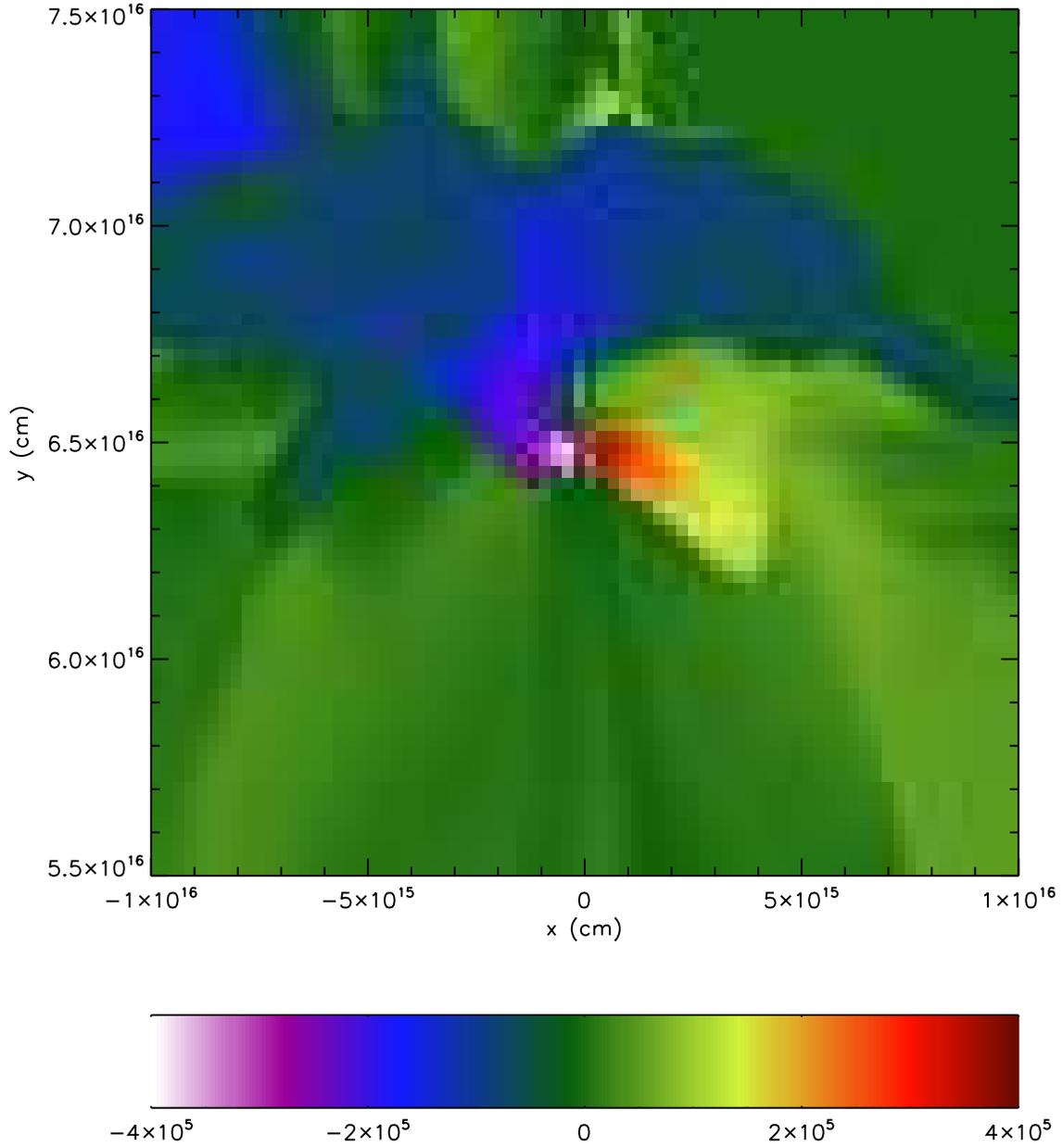}
\vspace{-1.0in}
\caption{Model 20-400-pr14 $z$ velocity, plotted as in Fig. 6, at 0.090 Myr.}
\end{figure}
\clearpage

\begin{figure}
\vspace{-1.0in}
\plotone{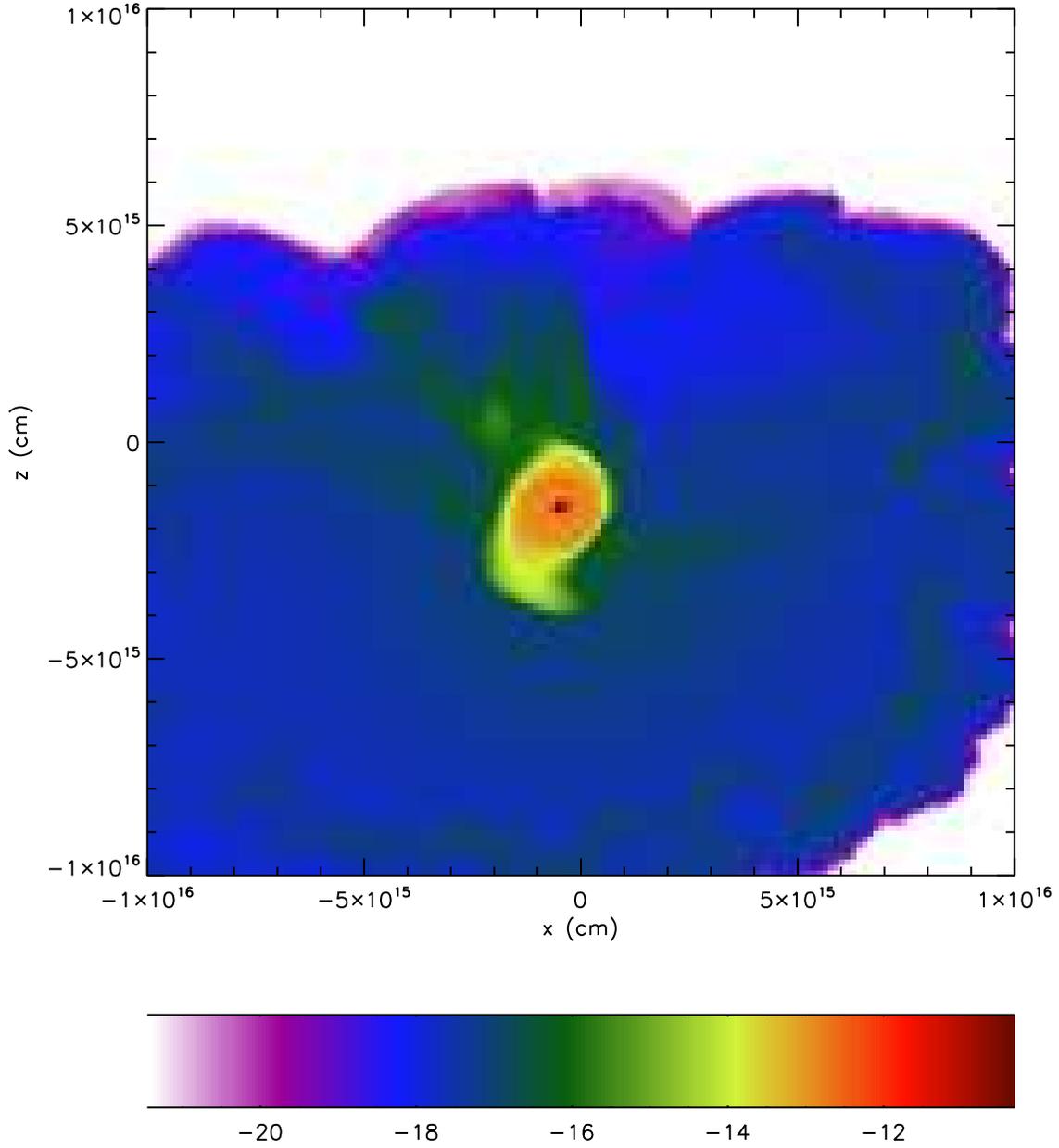}
\vspace{-1.0in}
\caption{Model 40-200-pr13 log density in the disk midplane 
($y = 2.24 \times 10^{16}$ cm) at 0.085 Myr.}
\end{figure}
\clearpage

\begin{figure}
\vspace{-1.0in}
\plotone{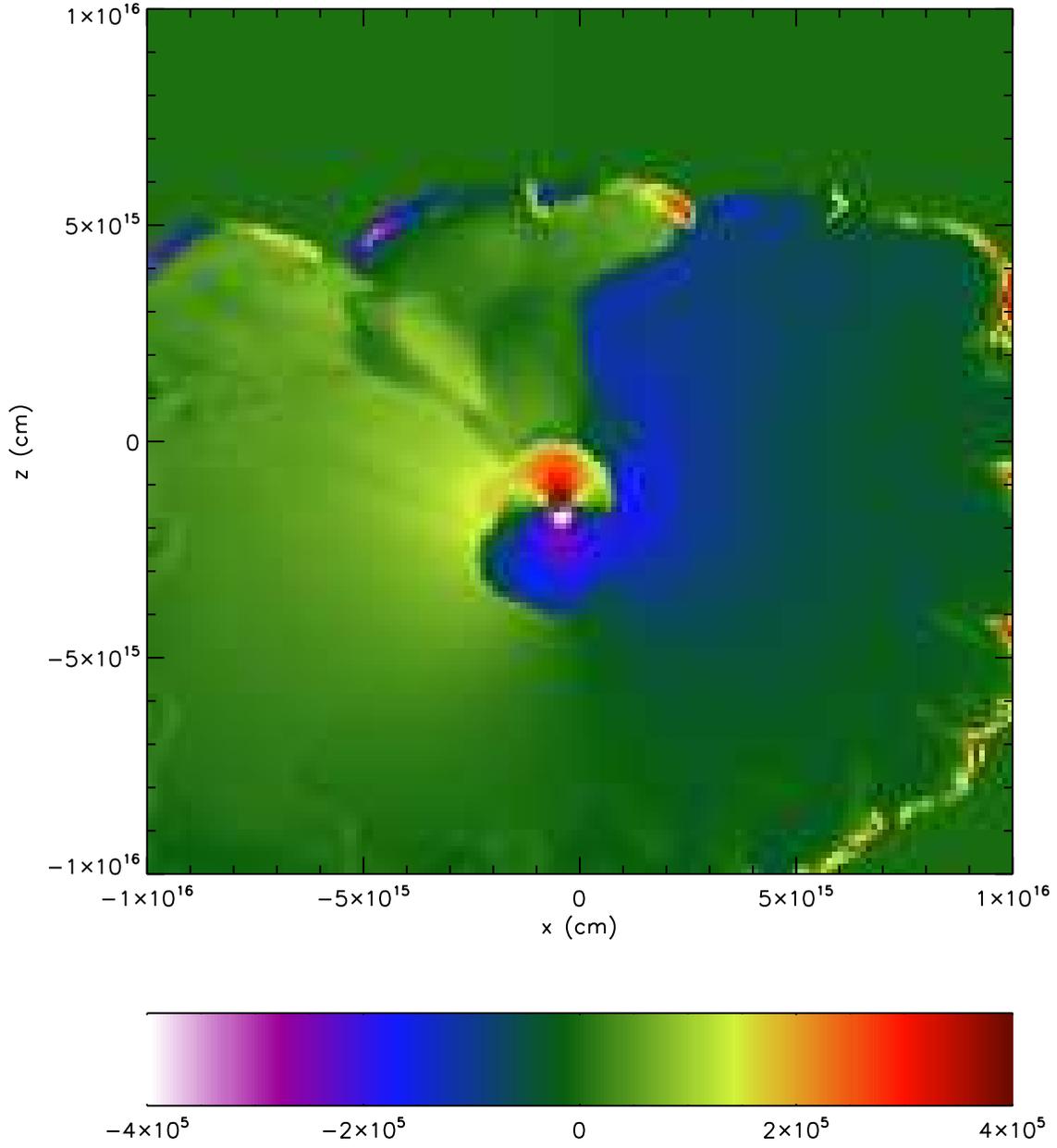}
\vspace{-1.0in}
\caption{Model 40-200-pr13 $x$ velocity in the disk midplane
($y = 2.24 \times 10^{16}$ cm) at 0.085 Myr.}
\end{figure}
\clearpage

\begin{figure}
\vspace{-1.0in}
\plotone{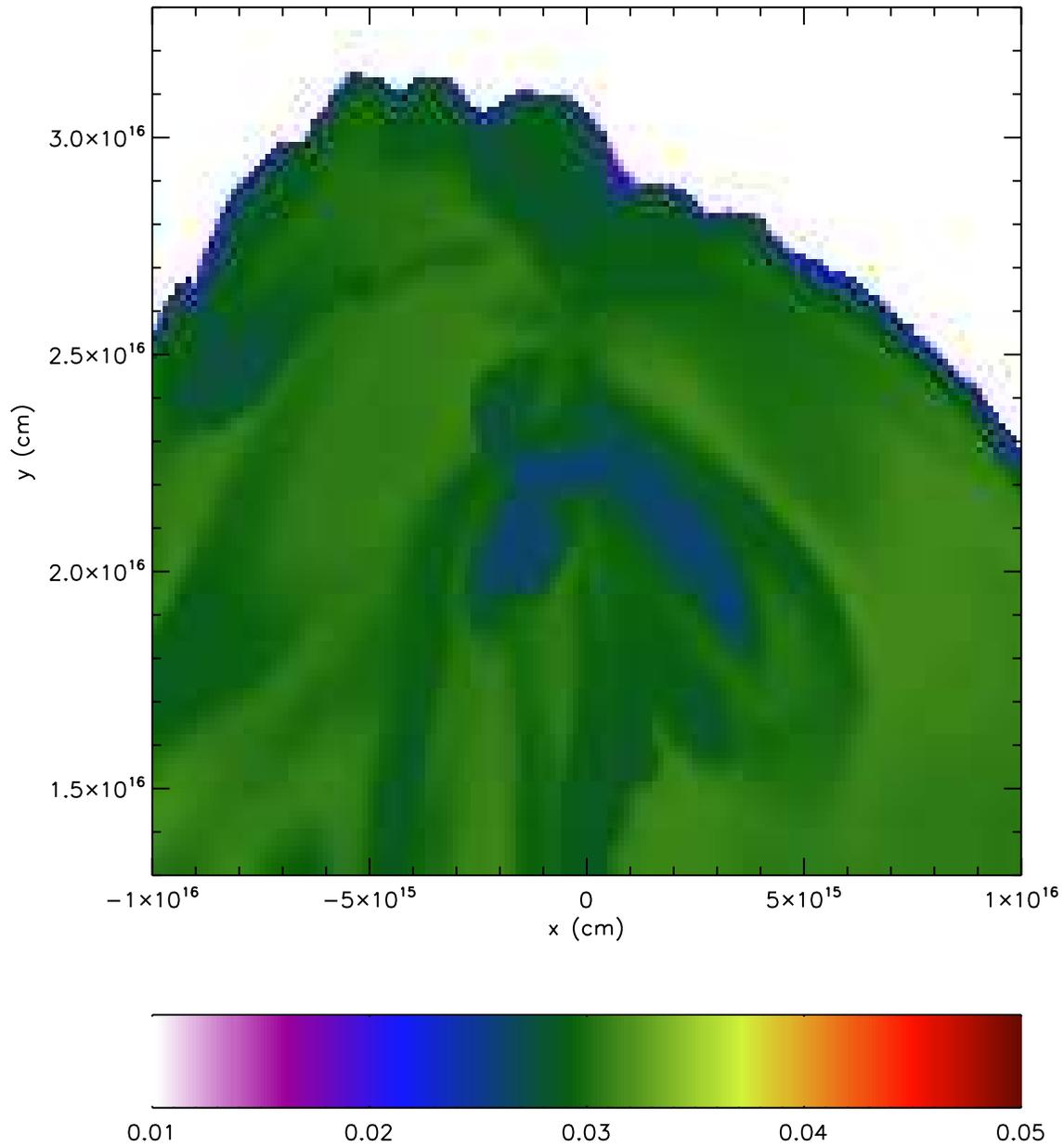}
\vspace{-1.0in}
\caption{Model 40-200-pr13 color density, plotted as in Fig. 1 at 0.085 Myr.}
\end{figure}
\clearpage

\begin{figure}
\vspace{-1.0in}
\plotone{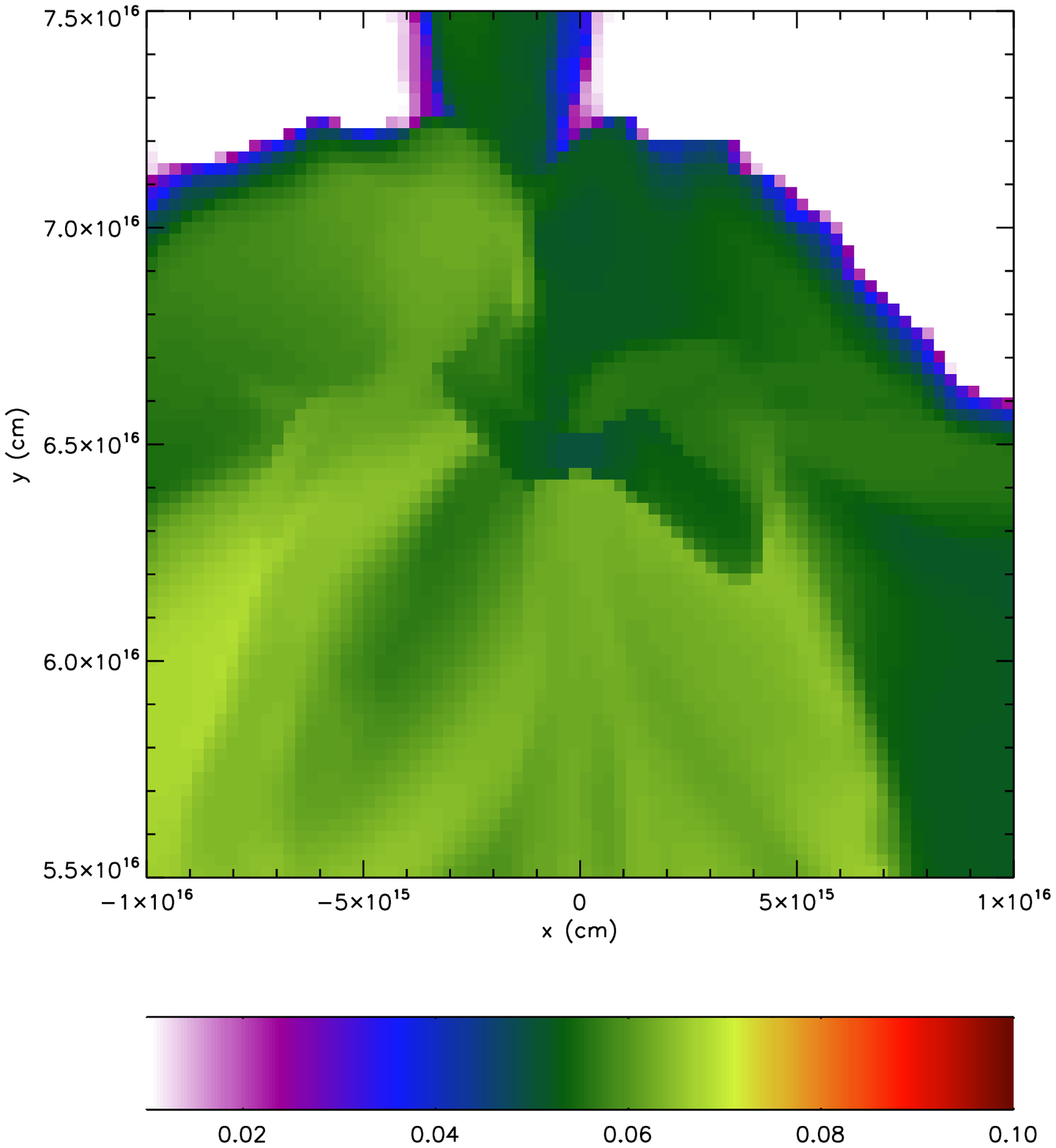}
\vspace{-1.0in}
\caption{Model 20-400-pr14 color density, plotted as in Fig. 6 at 0.090 Myr.}
\end{figure}
\clearpage


\begin{references}

\reference{r}
Adams, F. C., Fatuzzo, M., \& Holden, L. 2014, ApJ, 789, 86.

\reference{r}
Bermingham, K. R., Mezger, K., Desch, S. J., Scherer, E. E., \& 
Horstmann, M. 2014, Geochim. Cosmochim. Acta, 133, 463

\reference{r}
Blair, W. P., Sankrit, R., Raymond, J. C., \& Long, K. S. 1999, AJ, 118, 942

\reference{r}
Boss, A. P. 1995, ApJ, 439, 224

\reference{r}
Boss, A. P., \& Keiser, S. A. 2012, ApJL, 756, L9 

\reference{r}
Boss, A. P., \& Keiser, S. A. 2013, ApJ, 770, 51 

\reference{r}
Boss, A. P., \& Keiser, S. A. 2014, ApJ, 788, 20

\reference{r}
Boss, A. P., \& Yorke, H. A. 1995, ApJ,  439, L55

\reference{r}
Boss, A. P., Keiser, S. A., Ipatov, S. I., Myhill, E. A., \& Vanhala, 
H. A. T. 2010, ApJ, 708, 1268

\reference{r}
Cameron, A. G. W., \& Truran, J. W. 1977, Icarus, 30, 447

\reference{r}
Fryxell, B., Olson, K., Ricker, P., et al. 2000, ApJS, 131, 273

\reference{r}
Harsono, D., Jorgensen, J. K., van Dishoeck, E. F., et al. 2014, A\&A, 562, A77

\reference{r}
Kaufman, M. J., \& Neufeld, D. A. 1996, ApJ, 456, 611

\reference{r}
Li, S., Frank, A., \& Blackman, E. G.	 2014, MNRAS, 444, 2884

\reference{r}
Mishra, R. K., \& Chaussidon, M. 2014, EPSL, 398, 90

\reference{r}
Mishra, R. K., \& Goswami, J. N. 2014, Geochim. Cosmochim. Acta, 132, 440

\reference{r}
Neufeld, D. A., \& Kaufman, M. J. 1993, ApJ, 418, 263

\reference{r}
Ouellette, N., Desch, S. J., \& Hester, J. J. 2007, ApJ, 662, 1268

\reference{r}
Ouellette, N., Desch, S. J., \& Hester, J. J. 2010, ApJ, 711, 597

\reference{r}
Reach, W. T., Rho, J., \& Jarrett, T. H. 2005, ApJ, 618, 297

\reference{r}
Sashida, T., Oka, T., Tanaka, K., et al. 2013, ApJ, 774, 10

\reference{r}
Tachibana, S., Huss, G. R., Kita, N. T., Shimoda, G., \&
Morishita, Y. 2006, ApJ, 639, L87

\reference{r}
Takigawa, A., Miki, J., Tachibana, S., et al. 2008, ApJ, 688, 1382

\reference{r}
Tang, H., \& Dauphas, N. 2012, EPSL, 59, 248

\reference{r}
Tatischeff, V., Duprat, J. \& De S\'er\'eville, N. 2014, ApJ, 796, 124



\end{references}
\end{document}